\documentclass[10.7pt,draftcls,onecolumn,technote]{IEEEtran}
\usepackage[dvips]{epsfig}
\usepackage[dvips]{graphicx}
\usepackage{boxedminipage}
\usepackage{color}
\usepackage{amsfonts}
\usepackage{amssymb}
\usepackage{amsmath}
\usepackage{graphics,amsmath,amsfonts,amssymb,epsfig,latexsym,amsfonts,graphicx,amssymb}

\newcommand{\trace}{{\mbox{\textrm{Tr}}}}
\newcommand{\rank}{{\mbox{\textrm{Rank}}}}
\newcommand{\bfW}{{\mbox{$\mathbf{W}$}}}

\def\BibTeX{{\rm B\kern-.05em{\sc i\kern-.025em b}\kern-.08em
    T\kern-.1667em\lower.7ex\hbox{E}\kern-.125emX}}

\title{Lower Bounds Optimization for Coordinated
Linear Transmission Beamformer Design in Multicell Network Downlink}
\author{Mingyi Hong, Alfredo Garcia, J. Joaqu\'{i}n Escudero Garz\'{a}s, Ana Garc\'{i}a-Armada
\thanks{\small M. Hong and A. Garcia are with the Department of Systems and Information Engineering,
University of Virginia, USA (e-mail: \{mh4tk, ag7s\}@virginia.edu).
A. Garc\'{i}a-Armada is with the Department of Signal Theory and
Communications, University Carlos III of Madrid, Spain (e-mail:
agarcia@tsc.uc3m.es). J.J. Escudero-Garz\'{a}s is with the Group of
Signal Processing for Communications and Navigations (SPCOMNAV),
Universitat Autonoma de Barcelona, (e-mail:
josejoaquin.escudero@uab.cat)}}
\begin{document}
\maketitle
%\begin{abstract}
%The convergence properties of the Iterative water-filling (IWF)
%based algorithms (\cite{yu02a}, \cite{scutari08a}, \cite{wang08})
%have been derived in the {ideal situation} where the transmitters in
%the network are able to obtain the \emph{exact} value of the
%interference plus noise (IPN) experienced at the corresponding
%receivers \emph{in each iteration} of the algorithm. However, these
%algorithms are not robust because they diverge when there is {\it
%time-varying} estimation error of the IPN, a situation that arises
%in real communication system. In this correspondence, we propose an
%algorithm that possesses convergence guarantees in the presence of
%various forms of such time-varying error. Moreover, we also show by
%simulation that in scenarios where the interference is strong, the
%conventional IWF diverges while our proposed algorithm still
%converges.
%\end{abstract}
\vspace{-0.6cm}
\begin{abstract}
We consider the coordinated downlink beamforming problem in a
cellular network with the base stations (BSs) equipped with multiple
antennas, and with each user equipped with a single antenna. The BSs
cooperate in sharing their local interference information, and they
aim at maximizing the sum rate of the users in the network. A set of
new lower bounds (one bound for each BS) of the non-convex sum rate
is identified.  These bounds facilitate the development of a set of
algorithms that allow the BSs to update their beams by optimizing
their respective lower bounds. We show that when there is a single
user per-BS, the lower bound maximization problem can be solved
exactly with rank-1 solutions. In this case, the overall sum rate
maximization problem can be solved to a KKT point. Numerical results
show that the proposed algorithms achieve high system throughput
with reduced backhaul information exchange among the BSs.
\end{abstract}

\vspace{-0.5cm}
\section{Introduction}
Multiple Input - Multiple Output (MIMO) communications
\cite{foschini98} have been adopted in many recent wireless
standards, such as IEEE 802.16 \cite{ieee11} and 3GPP LTE %(3rd
%Generation Partnership Project Long Term Evolution)
\cite{3gpp09}, in the aim of boosting the data rates provided to the
customers. A promising solution to achieve spectrally-efficient
communications is the universal frequency reuse (UFR) scheme, in
which all cells operate on the same frequency channel. However, the
downlink capacity of the conventional cellular
systems with UFR is limited by inter-cell interference. %The classical approach to cope with interference in
%cellular systems has been frequency reuse with the inherent loss in
%spectral efficiency.
As a result, it is necessary to introduce coordination among the
base stations (BSs) so that they can jointly manage the
interferences in all cells to improve the system performance
\cite{gesbert07}.
%And such requirement leads to recent interest in techniques to
%manage interference in cellular systems with UFR, by introducing
Such coordination technique among the BSs in the downlink is also
known as network MIMO \cite{Karakayali06} or Coordinated multipoint
(CoMP) \cite{Sawahashi10}. Some other approaches in the literature
have exploited less complex linear schemes, such as Block
Diagonalization (BD) \cite{zhang09} or MMSE \cite{armanda11}. The
main drawback of all these systems is that they require channel
state information (CSI) and transmit data simultaneously known to
all cooperating BSs, with the cost of increased signal overhead.
Some recent approaches have been proposed to avoid CSI and data
sharing. Non-coherent joint processing \cite{sun11} does not require
cell-to-cell CSI exchange at the expense of higher processing cost
at the receivers with successive interference cancelation. In
\cite{bjornson10}, the authors analyze the case of distributed
cooperation where each BS has only local CSI.

In this correspondence we consider a cellular scenario with an
arbitrary number of multiantenna transmitters (the BSs) and
single-antenna receivers (the users). We focus on an {\it
intermediate approach} where the BSs optimize the downlink
throughput with only the CSI information. Since channel variations
are much slower than that of data, the amount and the
frequency of information exchange is greatly reduced. %Moreover, it
%is possible that each mobile station feeds back its CSI to all
%collaborating BSs (as suggested in \cite{Papadogiannis09}), so that
%the necessary infrastructural overheads and costs are even lower.

Unfortunately, the sum rate maximization problem is non-convex and
thus is difficult to solve efficiently. The authors of \cite{huh10}
propose to solve the single cell downlink rate maximization problem
first (with dirty paper coding (DPC) and zero-forcing (ZF)
precoding), and then impose interference limit to the users on the
cell edges. In this case, the interference limits to the users are
set in a rather heuristic fashion, and the BSs are not coordinating
their beamforming. References \cite{yu11} and \cite{venturino10} are
two recent works that propose heuristic algorithms that try to
provide solutions to similar problems by directly solving the
non-convex optimization problem.

In this correspondence we provide theoretical insights to the
coordinated downlink beamforming problem by identifying a set of
lower bounds (one bound per BS) of the non-convex system sum rate.
The benefits of such per-BS lower bounds are twofolds: 1) the
individual BSs can distributedly optimize their respective lower
bounds instead of jointly optimizing the original system sum rate to
approach a solution to the sum rate maximization problem; 2)
individual BSs can monitor the improvement of the total sum rate by
evaluating their respective lower bounds. Utilizing this set of
lower bounds, we propose algorithms for the BSs to coordinately
optimize their beams. In a special case where each cell has a single
user, each lower bound becomes {\it concave}, and we show that the
lower bound maximization problem can be solved exactly. This result
allows us to obtain a stationary solution of the original sum rate
maximization problem. In the general case with multiple users per
cell, we propose an algorithm that extend the Iterative Coordinated
Beamforming (ICBF) algorithm proposed in \cite{venturino10}, with
important difference that the BSs act sequentially instead of
simultaneously, and there is no ``inner iteration" needed. The
simulation results show that the proposed algorithms have similar
sum rate performance as the ICBF algorithm, while requiring
significantly less information exchange among the BSs in the
backhaul network.

The correspondence is organized as follows. In section
\ref{secSystemModel}, we give the system description, and provide a
general lower bound for each user. In section \ref{secSingleUser}
and  \ref{secMultipleUser}, we propose algorithms for the BSs to
compute their beamformers in different network configurations. In
section \ref{secSimulation}, we provide numerical results to
demonstrate the performance of the proposed algorithms. This
correspondence concludes in Section \ref{secConclusion}.

%In
%\cite{Bhagavatula11} an efficient feedback mechanism is designed for
%this CSI sharing.

{\it Notations}: For a symmetric matrix $\mathbf{X}$,
$\mathbf{X}\succeq 0$ signifies that $\mathbf{X}$ is positive
semi-definite. We use $\trace(\mathbf{X})$, $|\mathbf{X}|$,
$\mathbf{X}^H$, $\mathbf{X}^{\dag}$ and $\rank(\mathbf{X})$ to
denote the trace, the determinant, the hermitian, the pseudoinverse,
and the rank of a matrix, respectively. $[\mathbf{X}]_{i,i}$ denote
the $(i,i)$th element of the matrix $\mathbf{X}$. $\mathbf{I}_n$ is
used to denote a $n\times n$ identity matrix. We use
$[y,\mathbf{x}_{-i}]$ to denote a vector $\mathbf{x}$ with its
$i^{th}$ element replaced by $y$. We use $\mathbb{R}^{N\times M}$
and $\mathbb{C}^{N\times M}$ to denote the set of real and complex
$N\times M$ matrices; We use $\mathbb{S}^{N}$ and
$\mathbb{S}^{N}_{+}$ to denote the set of $N\times N$ hermitian and
hermitian semi-definite matrices, respectively. Define $M\oslash
t\triangleq \{(M+1)\mod t\}+1$ as an integer taking values from
$1,\cdots,M$.

\vspace{-0.3cm}
\section{Problem Formulation and System Model}
\label{secSystemModel}

We consider a multi-cell cellular network with a set
$\mathcal{M}\triangleq\{1,\cdots, M\}$ of base stations (BSs)/cells;
each BS is equipped with $K_m$ transmit antennas; each cell $m$ has
a set $\mathcal{N}_m$ of distinctive users; let $\mathcal{N}$ denote
the set of all users, and each user is equipped with a single
receive antenna. We use $(m,i)$ and $-(m,i)$ to denote the $i$th
user in $m$th cell and all the users except user $(m,i)$,
respectively. Without loss of generality, we assume that all the
cells have the same number of users, and all the BSs are equipped
with the same number of antennas:
$|\mathcal{N}_m|=N,~K_m=K,~\forall~m\in\mathcal{M}$. The signal
$\mathbf{x}_m\in\mathbb{C}^{K}$ transmitted by BS $m$ is
$\mathbf{x}_m=\sum_{i\in\mathcal{N}_m}\mathbf{w}_{m,i}b_{m,i}$,
where $b_{m,i}$ is the complex information symbol sent by  BS $m$ to
user $i\in\mathcal{N}_m$, using beam vector
$\mathbf{w}_{m,i}\in\mathbb{C}^{K}$. Assume $E[|b_{m,i}|^2]=1$, for
all $(m,i)$ and $E[b_{m,i}b^*_{q,j}]=0$, for all $(m,i)\ne (q,j)$.
Assume that each BS $m\in\mathcal{M}$ has a total transmission power
constraint:
$\sum_{i\in\mathcal{N}_m}||\mathbf{w}_{m,i}||^2\le\bar{p}_m$. Let
$\mathbf{h}_{q,m_i}\in\mathbb{C}^{K}$ denote the complex channel
between the $q$th BS and the $i$th user in $m$th cell. Let
$n_{m,i}\in\mathbb{C}$ denote the circularly-symmetric Gaussian
noise with variance ${c}_{m,i}$. The signal received by a user
$(m,i)$ can be expressed as
\begin{align}
y_{m,i}
&=\mathbf{h}^{H}_{m,i}\mathbf{w}_{m,i}b_{m,i}+\underbrace{\sum_{j\ne
i}\mathbf{h}^{H}_{m,m_i}\mathbf{w}_{m,j}b_{m,j}}_{\textrm{Intra-cell
Interference}}+\underbrace{\sum_{q\ne m,
j\in\mathcal{N}_q}\mathbf{h}^{H}_{q,m_i}\mathbf{w}_{q,j}b_{q,j}}_{\textrm{Inter-cell
Interference}}+n_{m,i}\label{eqReceivedSignal}.
\end{align}
 The rate achievable for user
$(m,i)$ is given by{\small
 \begin{align}
R_{m,i}(\mathbf{w}_{m,i},\mathbf{w}_{-(m,i)})&%=\log(1+SINR_{m,i})
\triangleq\log\left(1+\frac{\mathbf{w}^H_{m,i}\mathbf{H}_{m,m_i}\mathbf{w}_{m,i}}
{{c}_{m,i}+\sum_{(q,j)\ne
(m,i)}\mathbf{w}^H_{q,j}\mathbf{H}_{q,m_i}\mathbf{w}_{q,j}}\right)\label{eqRateScalar}\\
&=\log\left(1+\frac{\mathbf{h}^H_{m,m_i}\mathbf{W}_{m,i}\mathbf{h}_{m,m_i}}
{{c}_{m,i}+\sum_{(q,j)\ne
(m,i)}\mathbf{h}^H_{q,m_i}\mathbf{W}_{q,j}\mathbf{h}_{q,m_i}}\right)%\nonumber\\
%&=\log\left(1+\frac{\mathbf{h}^H_{m,i}\mathbf{W}_{m,i}\mathbf{h}_{m,i}}
%{I_{m,i}(\mathbf{W}_{-(m,i)})}\right)
\triangleq
R_{m,i}(\mathbf{W}_{m,i},\mathbf{W}_{-(m,i)})\label{eqRateMatrix}
 \end{align}}
where $\mathbf{W}_{m,i}\triangleq\mathbf{w}_{m,i}\mathbf{w}^H_{m,i}$
is the transmission covariance of user $(m,i)$, and
$\mathbf{H}_{m,m_i}\triangleq
\mathbf{h}_{m,m_i}\mathbf{h}^H_{m,m_i}$ is the channel matrix.
Clearly, $\mathbf{W}_{m,i}\succeq 0$ and
$\rank(\mathbf{W}_{m,i})=1$. Define the total interference plus
noise at user $(m,i)$ as{\small
\begin{align}
I_{m,i}(\mathbf{W}_{-(m,i)})&\triangleq {c}_{m,i}+\sum_{j\ne
i}\mathbf{h}^H_{m,m_i}\mathbf{W}_{m,j}\mathbf{h}_{m,m_i}+\sum_{q\ne
m, j\in\mathcal{N}_q}\mathbf{h}^H_{q,m_i}\mathbf{W}_{q,j}\mathbf{h}_{q,m_i}\nonumber\\
&={c}_{m,i}+\sum_{j\ne
i}\mathbf{w}^H_{m,j}\mathbf{H}_{m,m_i}\mathbf{w}_{m,j}+\sum_{q\ne m,
j\in\mathcal{N}_q}\mathbf{w}^H_{q,j}\mathbf{H}_{q,m_i}\mathbf{w}_{q,j}\triangleq
I_{m,i}(\mathbf{w}_{-(m,i)}).
\end{align}}
We assume that $I_{m,i}(\mathbf{W}_{-(m,i)})$ is perfectly known at
the user $(m,i)$ and the BSs $m$, but not the neighboring BSs. As
suggested by \cite{zhang09}, this interference plus noise term can
be estimated at each mobile user by various methods, and fed back to
its associated BS. Define the collection of matrices
$\mathbf{W}_m\triangleq\{\mathbf{W}_{m,i}\}_{i\in\mathcal{N}_m}$,
$\mathbf{W}_{-m}\triangleq\{\mathbf{W}_{q,j}\}_{j\in\mathcal{N}_q,
q\ne m}$, and
$\mathbf{W}\triangleq\{\mathbf{W}_m\}_{m\in\mathcal{M}}$, then the
sum rate of all users in cell $m$ can be expressed as: $
{R}_m(\mathbf{W}_{m},\mathbf{W}_{-m})\triangleq\sum_{i\in\mathcal{N}_m}
R_{m,i}(\mathbf{W}_{m,i}, \mathbf{W}_{-(m,i)})$. The sum rate of all
users in the network is $
{R}(\mathbf{W})\triangleq\sum_{q\in\mathcal{M}}
R_{q}(\mathbf{W}_{q},\mathbf{W}_{-q})$. We are interested in the
following non-concave sum rate maximization problem\footnote{This
problem can also be expressed in an equivalent vector form, with
$\{\mathbf{w}_m\}_{m\in\mathcal{M}}$ as design variables.}:{\small
\begin{align}
\max_{\mathbf{W}}& \quad R(\mathbf{W})\tag{SRM} \nonumber\\
{\textrm{s.t.}}&\quad\trace\left[\sum_{i\in\mathcal{N}_m}\mathbf{W}_{m,i}\right]\le
\bar{p}_m,~\forall~m\in\mathcal{M}\nonumber\\
& \quad \mathbf{W}_{m,i}\succeq 0,~\rank(\mathbf{W}_{m,i})\le
1,~~\forall~(m,i)\nonumber.
\end{align}}
We mention that all the following discussions are equally applicable
to the problem of {\it weighted} sum rate optimization, in which
there is a set of non-negative weights associated to the users'
rates in the objective. However, we mainly consider the (SRM)
problem for simplicity of presentation.

%\vspace{-0.3cm}
%\subsection{A Lower Bound for the System Sum
%Rate}\label{subLowerBound}
In order to approach the problem (SRM), we first establish some
useful results that characterize the users' rate
\eqref{eqRateMatrix}.
\newtheorem{P1}{Proposition}
\begin{P1}\label{propConvex}
{\it For all $(q,j)\ne(m,i)$, {\small
$R_{m,i}(\mathbf{W}_{m,i},\mathbf{W}_{-(m,i)})$} is a convex
function of {\small $\mathbf{W}_{q,j}$} on $\mathbb{S}^{K}_{+}$, and
a concave function of {\small $\mathbf{W}_{m,i}$} on
$\mathbb{S}^{K}_{+}$.}
\end{P1}
\begin{proof}
In order to show the convexity result, it is sufficient to prove
that whenever $\mathbf{D}\in \mathbb{S}^{K}$, $\mathbf{D}\ne
\mathbf{0}$ and $\mathbf{W}_{q,j}+t\mathbf{D}\succeq 0$, the
following function is convex in $t$ \cite[Chapter 3]{cover05}
{\small
\begin{align}
R_{m,i}(t)&\triangleq\log\left(1+\frac{\mathbf{h}^H_{m,m_i}\mathbf{W}_{m,i}\mathbf{h}_{m,m_i}}
{{c}_{m,i}+\sum_{(p,l)\ne (q,j), (p,l)\ne
(m,i)}\mathbf{h}^H_{p,m_i}\mathbf{W}_{p,l}\mathbf{h}_{p,m_i}+
\mathbf{h}^H_{q,m_i}(\mathbf{W}_{q,j}+t\mathbf{D})\mathbf{h}_{q,m_i}}\right).
\end{align}}
Let us simplify the expression a bit by defining the constant
$c=\mathbf{h}^H_{m,m_i}\mathbf{W}_{m,i}\mathbf{h}_{m,m_i}\ge0$ (note
that $\mathbf{W}_{m,i}\succeq 0$). The first and the second
derivatives of $R_{m,i}(t)$ w.r.t. $t$ can be expressed as{\small
\begin{align}
\frac{d R_{m,i}(t)}{d t}
&=-\frac{1/\ln(2)}{\left(I_{m,i}(\mathbf{W}_{-(m,i)})+
t\mathbf{h}^H_{q,m_i}\mathbf{D}\mathbf{h}_{q,m_i}+c\right)}\frac{c
\mathbf{h}^H_{q,m_i}\mathbf{D}\mathbf{h}_{q,m_i}}{\left(I_{m,i}(\mathbf{W}_{-(m,i)})+
t\mathbf{h}^H_{q,m_i}\mathbf{D}\mathbf{h}_{q,m_i}\right)}.\label{eqFirstDerivativeT}\\
\frac{d^2 R_{m,i}(t)}{d
t^2}&=\frac{1/\ln(2)}{(I_{m,i}(\mathbf{W}_{-(m,i)})+
t\mathbf{h}^H_{q,m_i}\mathbf{D}\mathbf{h}_{q,m_i}+c)^2}\frac{c
(\mathbf{h}^H_{q,m_i}\mathbf{D}\mathbf{h}_{q,m_i})^2}{I_{m,i}(\mathbf{W}_{-(m,i)})+
t\mathbf{h}^H_{q,m_i}\mathbf{D}\mathbf{h}_{q,m_i}}\nonumber\\
&+
\frac{1/\ln(2)}{I_{m,i}(\mathbf{W}_{-(m,i)})+t\mathbf{h}^H_{q,m_i}\mathbf{D}\mathbf{h}_{q,m_i}+c}\frac{c
(\mathbf{h}^H_{q,m_i}\mathbf{D}\mathbf{h}_{q,m_i})^2}{(I_{m,i}(\mathbf{W}_{-(m,i)})+
t\mathbf{h}^H_{q,m_i}\mathbf{D}\mathbf{h}_{q,m_i})^2}.
\end{align}}
Clearly  $I_{m,i}(\mathbf{W}_{-(m,i)})+
t\mathbf{h}^H_{q,m_i}\mathbf{D}\mathbf{h}_{q,m_i}> 0$ for all
$\mathbf{W}_{q,j}+t\mathbf{D}\succeq 0$. We also have that
$\mathbf{h}^H_{q,m_i}\mathbf{D}\mathbf{h}_{q,m_i}$ is real and
$(\mathbf{h}^H_{q,m_i}\mathbf{D}\mathbf{h}_{q,m_i})^2\ge 0$, due to
the assumption that $\mathbf{D}\in\mathbb{S}^K$, and the subsequent
implication that
$(\mathbf{h}^H_{q,m_i}\mathbf{D}\mathbf{h}_{q,m_i})^{H}=\mathbf{h}^H_{q,m_i}\mathbf{D}\mathbf{h}_{q,m_i}$.
We conclude that whenever $\mathbf{D}\in\mathbb{S}^{K}$ and
$\mathbf{W}_{q,j}+t\mathbf{D}\succeq 0$, $\frac{d^2 R_{m,i}(t)}{d
t^2}\ge 0$, which implies that
$R_{m,i}(\mathbf{W}_{m,i},\mathbf{W}_{-(m,i)})$ is convex in
$\mathbf{W}_{q,j}$ for all $(q,j)\ne (m,i)$.

The fact that $R_{m,i}(\mathbf{W}_{m,i},\mathbf{W}_{-(m,i)})$ is
concave in $\mathbf{W}_{m,i}$ can be shown similarly as above.
\end{proof}

Note that the above  property is only true in the space of
covariance matrix $\mathbf{W}_m$, but not in the transmit beamformer
space $\mathbf{w}_m$.  This convex-concave property of the
individual users' transmission rate is instrumental in deriving a
set of lower bounds for the system sum rate. For a particular user
$(m,i)$, the system sum rate $R(\mathbf{W})$ can be expressed
as{\small
\begin{align}
R(\mathbf{W})=\underbrace{R_{m,i}(\mathbf{W}_{m,i},\mathbf{W}_{-(m,i)})}_{\textrm{
concave in $\mathbf{W}_{m,i}$
}}+\underbrace{\sum_{(q,j)\ne(m,i)}R_{q,j}(\mathbf{W}_{m,i},\mathbf{W}_{-(m,i)})}_{\textrm{
convex in $\mathbf{W}_{m,i}$}}.\label{eqConvexConcave}
\end{align}}
Defined {\small
$R_{-(m,i)}\left({\mathbf{W}}\right)\triangleq\sum_{(q,j)\ne(m,i)}R_{q,j}\left({\mathbf{W}}\right)$}.
We can find a lower bound for  $R(\mathbf{W})$ by linearizing the
 $R_{-(m,i)}\left({\mathbf{W}}\right)$ with respect
to $\mathbf{W}_{m,i}$ around a fixed $\widehat{\mathbf{W}}$.
Utilizing the fact that $R_{-(m,i)}\left({\mathbf{W}}\right)$ is
convex in $\mathbf{W}_{m,i}$, we obtain {\small
\begin{align}
\sum_{(q,j)\ne(m,i)}R_{q,j}\left(\mathbf{W}_{m,i},\widehat{\mathbf{W}}_{-(m,i)}\right)
&\ge R_{-(m,i)}\left(\widehat{\mathbf{W}}\right)-
\sum_{(q,j)\ne(m,i)}\trace\left[T_{q,j}\left(\widehat{\mathbf{W}}\right)\mathbf{H}_{m,q_j}(\mathbf{W}_{m,i}-
\widehat{\mathbf{W}}_{m,i})\right]\label{eqApproximation}\\
\textrm{with}\quad\quad\quad
T_{q,j}\left(\widehat{\mathbf{W}}\right)&\triangleq
\frac{1/\ln(2)}{I_{q,j}(\widehat{\mathbf{W}}_{-(q,j)})+
\mathbf{h}^H_{q,j}\widehat{\mathbf{W}}_{q,j}\mathbf{h}_{q,j}}
\frac{\mathbf{h}^H_{q,q_j}\widehat{\mathbf{W}}_{q,j}\mathbf{h}_{q,q_j}}{I_{q,j}(\widehat{\mathbf{W}}_{-(q,j)})}\ge
0 \label{eqTax}.
\end{align}}
Let us define a concave function of $\mathbf{W}_{m,i}$ {\small
\begin{align}
U_{m,i}(\mathbf{W}_{m,i},\widehat{\mathbf{W}}_{-(m,i)})&\triangleq
R_{m,i}(\mathbf{W}_{m,i},\widehat{\mathbf{W}}_{-(m,i)})
+R_{-(m,i)}\left(\widehat{\mathbf{W}}\right)-
\sum_{(q,j)\ne(m,i)}\trace\left[T_{q,j}\left(\widehat{\mathbf{W}}\right)\mathbf{H}_{m,q_j}(\mathbf{W}_{m,i}-
\widehat{\mathbf{W}}_{m,i})\right].\nonumber
\end{align}}
Then from \eqref{eqConvexConcave}, \eqref{eqApproximation} and the
definition of $U_{m,i}(.)$, we must have{\small
\begin{align}
U_{m,i}(\mathbf{W}_{m,i},\widehat{\mathbf{W}}_{-(m,i)})\le
R(\mathbf{W}_{m,i},\widehat{\mathbf{W}}_{-(m,i)}),~\forall \
\mathbf{W}_{m,i}\succeq 0 \label{eqLowerBound}
\end{align}}
where the equality is achieved when
$\mathbf{W}_{m,i}=\widehat{\mathbf{W}}_{m,i}$. We refer to this
lower bound as the ``per-user" lower bound, as it is defined w.r.t.
each user $(m,i)$. Such lower bound is useful, because if we can
find a $\mathbf{W}^*_{m,i}$ that satisfies {\small$
U_{m,i}({\mathbf{W}}^*_{m,i},\widehat{\mathbf{W}}_{-(m,i)})>
U_{m,i}(\widehat{\mathbf{W}}_{m,i},\widehat{\mathbf{W}}_{-(m,i)})$},
then the system sum rate must increase, as{\small
\begin{align}
R(\mathbf{W}^*_{m,i},\widehat{\mathbf{W}}_{-(m,i)})\ge
U_{m,i}(\mathbf{W}^*_{m,i},\widehat{\mathbf{W}}_{-(m,i)} )>
U_{m,i}(\widehat{\mathbf{W}}_{m,i},\widehat{\mathbf{W}}_{-(m,i)}
)=R(\widehat{\mathbf{W}}_{m,i},\widehat{\mathbf{W}}_{-(m,i)}).\label{eqRateIncreasePerUser}
\end{align}}

\vspace{-0.5cm}
\section{Multi-cell Network with Single User In Each
Cell}\label{secSingleUser}

We first consider an important scenario in which each BS transmits
to a single user. This scenario may arise in a heterogeneous network
when each BS transmits to a relay in its cell. As there is a single
user in each cell, we simplify the notation by using $U_m(.)$,
$T_q(.)$, $I_m(.)$ instead of $U_{m,i}(.)$,  $T_{q,i}(.)$ and
$I_{m,i}(.)$, respectively. We use $\mathbf{W}_m$ to denote the
covariance of BS $m$ to its user; we use $\mathbf{H}_{m,q}$ to
denote the channel between BS $m$ to the user in the cell of BS $q$.
Notice that the per-user bound identified in Section
\ref{secSystemModel} becomes {\it per-BS} bound, as each BS has a
single user in this scenario. For simplicity, define $\sum_{q\ne
m}T_{q}\left(\widehat{\mathbf{W}}\right)\mathbf{H}_{m,q}=\mathbf{A}_m\succeq
0$, then the per-BS bound can be expressed as: {\small \begin{align}
 U_{m}(\mathbf{W}_{m},\widehat{\mathbf{W}}_{-m})\triangleq
R_{m}(\mathbf{W}_{m},\widehat{\mathbf{W}}_{-m})
+R_{-m}\left(\widehat{\mathbf{W}}\right)-
\trace\left[\mathbf{A}_m(\mathbf{W}_{m}-
\widehat{\mathbf{W}}_{m})\right].
\end{align}}
Define the feasible set for BS $m$ as
$\mathcal{F}_m\triangleq\{\mathbf{W}_m:
\trace\left[\mathbf{W}_{m}\right]\le \bar{p}_m,~
\mathbf{W}_{m}\succeq 0,~\rank(\mathbf{W}_{m})\le 1\}$. The idea is
to let the BSs take turns to optimize their respective lower bounds
$\{U_m(.)\}$. Assuming other BSs' transmissions are fixed as
$\widehat{\mathbf{W}}_{-m}$, the Lower Bound Maximization problem
(LBM) for BS $m$ is{\small
\begin{align}
&\max_{\mathbf{W}_m\in\mathcal{F}_m}U_m(\mathbf{W}_m,
\widehat{\mathbf{W}}_{-m})\tag{LBM}.
\end{align}}
Notice that after relaxing the rank constraint, the problem (LBM) is
a concave problem in the variable $\mathbf{W}_m$. In the sequel, we
will refer to the problem (LBM) {\it without} the rank constraint as
(R-LBM), and define its feasible set as $\mathcal{F}^{R}_m\triangleq
\{\mathbf{W}_m: \trace\left[\mathbf{W}_{m}\right]\le \bar{p}_m,~
\mathbf{W}_{m}\succeq 0\}$.

The problem (R-LBM) is a concave determinant maximization (MAXDET)
problem \cite{vandenberghe98}, and can be solved efficiently using
convex program/SDP solvers such as CVX \cite{cvx}. However, in
practice such general purpose solver may still induce heavy
computational burden. Moreover, the resulting optimal solution of
the relaxed problem may have rank greater than one. Fortunately,
these difficulties can be resolved. We have found an explicit
construction that generates a rank-1 solution of the problem (R-LBM)
(hence the optimal solution of problem (LBM)). The rank reduction
problem of downlink beamforming has been recently studied in
\cite{huang10}, \cite{wiesel08} and \cite{huh10}. However the
algorithms proposed in those works cannot be directly used to obtain
a rank-1 solution to (LBM): reference \cite{huang10} considers
problems with linear objective functions; references \cite{huh10}
and \cite{wiesel08} consider the relaxation of the MAXDET problem
{\it without} the linear penalty terms \footnote{With linear penalty
in the form of ${\small -\trace\left[\mathbf{A}_m(\mathbf{W}_{m}-
\widehat{\mathbf{W}}_{m,q})\right]}$, equation (43) is no longer
equivalent to equation (44) in \cite{wiesel08}.}.

Removing all the terms in the objective of (R-LBM) that are not
related to $\bfW_m$, we can write the partial Lagrangian of the
problem (R-LBM) as{\small
\begin{align}
L(\mathbf{W}_m,\mu_m)=\log\bigg|\mathbf{I}+\mathbf{W}_m\mathbf{H}_{m,m}\frac{1}{I_{m}(\widehat{\bfW}_{-m})}
\bigg|-\mbox{Tr}[(\mathbf{A}_m+\mu_n\mathbf{I})\mathbf{W}_m]+\mu_m\bar{p}_m
\end{align}}
where $\mu_m\ge0$ is the Lagrangian multiplier associated with the
power constraint. Notice the fact that $\mathbf{A}_m\succeq 0$, then
for any $\mu_m>0$, we can perform the Cholesky decomposition
$\mathbf{A}_m+\mu_m\mathbf{I}=\mathbf{L}^H\mathbf{L}$, which results
in
$\mbox{Tr}[(\mathbf{A}_m+\mu_m\mathbf{I})\mathbf{W}_m]=\mbox{Tr}[\mathbf{L}\mathbf{W}_m\mathbf{L}^H]$.
Define
$\bar{\mathbf{W}}_m(\mu_m)=\mathbf{L}\mathbf{W}_m\mathbf{L}^H$, %and
%use the identity
%$\log(1+\mathbf{a}^{H}\mathbf{b})=\log|\mathbf{I}+\mathbf{b}\mathbf{a}^H|$
%for two vectors $\mathbf{a},\ \mathbf{b}$,
we have{\small
\begin{align}
L(\mathbf{W}_m,\mu_m)&=\log\bigg|\mathbf{I}+\mathbf{L}^{-1}\bar{\mathbf{W}}_m(\mu_m)
\mathbf{L}^{-H}\mathbf{H}_{m,m}\frac{1}{I_{m}(\widehat{\bfW}_{-m})}\bigg|-
\mbox{Tr}[\bar{\mathbf{W}}_m(\mu_m)]+\mu_m\bar{p}_m\nonumber\\
&\stackrel{(a)}=\log\left|\mathbf{I}+\bar{\mathbf{W}}_m(\mu_m)\mathbf{V}\mathbf{\Delta}\mathbf{V}^H\right|-
\mbox{Tr}[\bar{\mathbf{W}}_m(\mu_m)]+\mu_m\bar{p}_m\nonumber\\
&\stackrel{(b)}=\log\left|\mathbf{I}+\widehat{\mathbf{W}}_m(\mu_m)\mathbf{\Delta}\right|-
\mbox{Tr}[\mathbf{V}\widehat{\mathbf{W}}_m(\mu_m)\mathbf{V}^H]+\mu_m\bar{p}_m\nonumber\\
&=\log\left|\mathbf{I}+\widehat{\mathbf{W}}_m(\mu_m)\mathbf{\Delta}\right|-
\mbox{Tr}[\widehat{\mathbf{W}}_m(\mu_m)]+\mu_m\bar{p}_m=L(\widehat{\mathbf{W}}_m(\mu_m))
\end{align}}
where in $(a)$ we have used the eigendecomposition:
$\mathbf{L}^{-H}\mathbf{H}_{m,m}\mathbf{L}^{-1}\frac{1}{I_{m}(\widehat{\bfW}_{-m})}
=\mathbf{V}\mathbf{\Delta}\mathbf{V}^H$; in $(b)$ we have defined
$\widehat{\mathbf{W}}_m(\mu_m)=\mathbf{V}^H\bar{\mathbf{W}}_m(\mu_m)\mathbf{V}$.
Let $\widehat{\mathbf{W}}_m^*(\mu_m)$ denote an optimal solution to
the problem $\max_{\widehat{\mathbf{W}}_m(\mu_m)\succeq 0}
L(\widehat{\mathbf{W}}_m(\mu_m))$.

We claim that there must exist a $\widehat{\mathbf{W}}^*_m(\mu_m)$
that is {\it diagonal}. Note that $\rank(\mathbf{H}_{m,m})=1$
implies $\rank(\mathbf{\Delta})\le 1$. Thus
$\widehat{\mathbf{W}}^*_m(\mu_m)\mathbf{\Delta}$ has at most a {\it
single column}. This implies that we can remove the off diagonal
elements of
$\mathbf{I}+\widehat{\mathbf{W}}^*_m(\mu_m)\mathbf{\Delta}$ without
changing the values of
$\left|\mathbf{I}+\widehat{\mathbf{W}}^*_m(\mu_m)\mathbf{\Delta}\right|$.
Consequently, for any given $\widehat{\mathbf{W}}^*_m(\mu_m)$, we
can construct a diagonal optimal solution
$\widehat{\mathbf{W}}^{*,D}_m(\mu_m)$ by removing all its off
diagonal elements. This operation removes all the off diagonal
elements of
$\mathbf{I}+\widehat{\mathbf{W}}^*_m(\mu_m)\mathbf{\Delta}$, and it
does not change either
$\left|\mathbf{I}+\widehat{\mathbf{W}}^*_m(\mu_m)\mathbf{\Delta}\right|$
or $\textrm{Tr}[\widehat{\mathbf{W}}^*_m(\mu_m)]$. Consequently
$\widehat{\mathbf{W}}^{*,D}_m(\mu_m)$ is also optimal. When
restricting $\widehat{\mathbf{W}}^{*}_m(\mu_m)$ to be diagonal, we
can find its closed-form expression{\small
\begin{align}
[\widehat{\mathbf{W}}_m^*(\mu_m)]_{i,i}=\left[\frac{[\mathbf{\Delta}]_{i,i}-1}{[\Delta]_{i,i}}\right]^+,\textrm{if
$[\mathbf{\Delta}]_{i,i}\ne 0$;}\quad
[\widehat{\mathbf{W}}_m^*(\mu_m)]_{i,i}=0,\
\textrm{otherwise},\label{eqWCompute}
\end{align}}
where $[x]^+=\max\{0,x\}$. Then we can obtain
${\mathbf{W}}^*_m(\mu_m)=\mathbf{L}^{-1}\mathbf{V}\widehat{\mathbf{W}}_m^{*}(\mu_m)\mathbf{V}^H\mathbf{L}^{-H}$.
Combining the fact that $\rank(\mathbf{\Delta})\le 1$ with
\eqref{eqWCompute} we conclude
$\rank(\widehat{\mathbf{W}}^*(\mu_m))\le 1$, and consequently
$\rank({\mathbf{W}}^*_m(\mu_m))\le 1$, {\it for any $\mu_m>0$}.

It is relatively straightforward to show that
$\mbox{Tr}[\mathbf{W}_m^*(\mu_m)]$ is strictly decreasing with
respect to $\mu_m$. Consequently if the optimal multiplier
$\mu_m^*>0$, then a bisection method can be used to find $\mu^*_m$
that satisfies the feasibility conditions
$\mbox{Tr}[\mathbf{W}_m^*(\mu^*_m)]\le \bar{p}_m$. Furthermore, we
can also show that when $\mu^*_m=0$, $\mathbf{A}_m$ must have full
rank. In this case, we can find the Cholesky decomposition
$\mathbf{A}_m=\mathbf{L}\mathbf{L}^H$, and the above construction
can still be used to directly obtain $\mathbf{W}^*_m(0)$ (without
bisection), that satisfy $\rank(\mathbf{W}_m^*(0))\le 1$.

In conclusion, for any $\mu_m^*\ge 0$, we obtain
$\rank({\mathbf{W}}^*_m(\mu^*_m))\le 1$. Table
\ref{tableUtilityMaximization} summarizes the above procedure.
\begin{table}[htb]
\begin{center} \vspace{-0.1cm} \caption{ The Optimization
of (LBM)} \label{tableUtilityMaximization} {
\begin{tabular}{|l|}
\hline S1) Choose $\mu^u_m$ and $\mu^l_m$ such that $\mu_m^*$ lies
in
$[\mu^l_m,~\mu^u_m]$.\\
S2) Let $\mu^{mid}_m=(\mu^l_m+\mu^u_m)/2$. Compute decomposition:\\
~~~~~$\mathbf{L}^H\mathbf{L}=\mathbf{A}_m+\mu^{mid}_m\mathbf{I}$\\
~~~~~$\mathbf{V}\Delta\mathbf{V}^H=
\mathbf{L}^{-H}\mathbf{H}_{m,m}\mathbf{L}^{-1}\frac{1}{I_{m}(\widehat{\bfW}_{-m})}$.\\
S3) Compute $\widehat{\mathbf{W}}^*_m(\mu^{mid}_m)$ by \eqref{eqWCompute}.\\
S4) Compute
$\mathbf{W}_m^*(\mu^{mid}_m)=\mathbf{L}^{-1}\mathbf{V}\widehat{\mathbf{W}}_m^{*}(\mu^{mid}_m)
\mathbf{V}^H\mathbf{L}^{-H}$.\\
S5) If $\mbox{Tr}(\mathbf{W}_m^*(\mu^{mid}_m))>\bar{p}_m$, let
$\mu^{l}_m=\mu^{mid}_m$; otherwise let $\mu^{u}_m=\mu^{mid}_m$.\\
S6) If $|\mbox{Tr}(\mathbf{W}_m^*(\mu^{mid}_m))-\bar{p}_m|<\epsilon$
or $|\mu^{u}_m-\mu^{l}_m|<\epsilon$, stop; otherwise go to S2).
\\
 \hline
\end{tabular}}
\end{center}
\vspace{-0.5cm}
\end{table}

In the following, we identify a special structure of the problem
(R-LBM) that allows it to admit a rank-1 solution. To this end, we
tailor the rank reduction procedure (abbreviated as RRP) proposed in
\cite{huang10} to fit our problem \footnote{Note that the RRP
procedure in \cite{huang10} cannot be directly applied to our
problem.  This is because in \cite{huang10}, the RRP is used to
identify rank-1 solution of semidefinite programs with linear
objective and constraints. Our problem is different in that the
objective function is of a logdet form. }. Assume that using
standard optimization package we obtain an optimal solution
$\widetilde{\mathbf{W}}^*_m$ to the convex problem (R-LBM), with
$\rank(\widetilde{\mathbf{W}}^*_m)=r>1$. Let
$\widetilde{\mathbf{W}}^{(1)}_m=\widetilde{\mathbf{W}}^*_m$, and let
$r^{(1)}=r$. At iteration $t$ of the the RRP, we perform a eigen
decomposition
$\widetilde{\mathbf{W}}^{(t)}_m=\mathbf{V}^{(t)}{\mathbf{V}^{(t)}}^H$,
where $\mathbf{V}^{(t)}\in\mathbb{C}^{K\times r^{(t)}}$. If
$R^{(t)}>1$, find $\mathbf{D}^{(t)}\in\mathbb{S}^{r^{(t)}}$ such
that the following three conditions are satisfied
\begin{align}
&\trace(\mathbf{D}^{(t)}{\mathbf{V}^{(t)}}^H\mathbf{H}_{m,m}\mathbf{V}^{(t)})=0\label{eqDH}\\
&\trace(\mathbf{D}^{(t)}{\mathbf{V}^{(t)}}^H\mathbf{A}_m\mathbf{V}^{(t)})=0\label{eqDA}\\
&\trace(\mathbf{D}^{(t)}{\mathbf{V}^{(t)}}^H\mathbf{V}^{(t)})=0\label{eqDI}.
\end{align}
If such $\mathbf{D}^{(t)}$ cannot be found, exit. Otherwise, let
$\lambda(\mathbf{D}^{(t)})$ be the eigenvalue of $\mathbf{D}^{(t)}$
with the largest absolute value, and construct
$\widetilde{\mathbf{W}}^{(t+1)}_m=\mathbf{V}^{(t)}(\mathbf{I}_r-\frac{1}{\lambda(\mathbf{D}^{(t)})}\mathbf{D}^{(t)})
{\mathbf{V}^{(t)}}^H\succeq 0$. Clearly,
$\rank(\mathbf{I}_r-\frac{1}{\lambda(\mathbf{D}^{(t)})})\le
r^{(t)}-1$, as a result, $\rank(\widetilde{\mathbf{W}}^{(t+1)}_m)\le
\rank(\widetilde{\mathbf{W}}^{(t)}_m)-1$, i.e., the rank has been
reduced by at least one. Utilizing \eqref{eqDH}--\eqref{eqDI}, we
obtain{\small
\begin{align}
&\mathbf{h}^H_{m,m}\widetilde{\mathbf{W}}^{(t+1)}_m\mathbf{h}_{m,m}
=\trace[\mathbf{H}_{m,m}\widetilde{\mathbf{W}}^{(t+1)}_m]\nonumber
=\trace\left[\mathbf{H}_{m,m}\mathbf{V}^{(t)}(\mathbf{I}_r-\frac{1}{\lambda(\mathbf{D}^{(t)})}\mathbf{D}^{(t)})
{\mathbf{V}^{(t)}}^H\right]\nonumber\\
&\quad\quad=\trace[\mathbf{H}_{m,m}\widetilde{\mathbf{W}}^{(t)}_m]
=\mathbf{h}^H_{m,m}\widetilde{\mathbf{W}}^{(t)}_m\mathbf{h}^{H}_{m,m}\label{eqHCancel}\\
&\trace[\mathbf{A}_m\widetilde{\mathbf{W}}^{(t+1)}_m]
=\trace\left[\mathbf{A}_m\mathbf{V}^{(t)}(\mathbf{I}_r-\frac{1}{\lambda(\mathbf{D}^{(t)})}\mathbf{D}^{(t)})
{\mathbf{V}^{(t)}}^H\right]
=\trace[\mathbf{A}_m\widetilde{\mathbf{W}}^{(t)}_m]\label{eqACancel}\\
&\trace[\widetilde{\mathbf{W}}^{(t+1)}_m]
=\trace\left[\mathbf{V}^{(t)}(\mathbf{I}_r-\frac{1}{\lambda(\mathbf{D}^{(t)})}\mathbf{D}^{(t)})
{\mathbf{V}^{(t)}}^H\right]
=\trace[\widetilde{\mathbf{W}}^{(t)}_m]\label{eqWCancel}.
\end{align}}
Equation \eqref{eqHCancel} and \eqref{eqACancel} ensure that the
objective value of (R-LBM) does not change, i.e.,
$U_m(\widetilde{\mathbf{W}}^{(t+1)}_m,
\widehat{\bfW}_{-m})=U_m(\widetilde{\mathbf{W}}^{(t)}_m,
\widehat{\bfW}_{-m})$. Equation \eqref{eqWCancel} ensures
$\trace[\widetilde{\mathbf{W}}^{(t+1)}_m]=\trace[\widetilde{\mathbf{W}}^{(t)}_m]\le
\bar{p}_m$. Combined with the fact that
$\widetilde{\mathbf{W}}^{(t+1)}_m\succeq 0$, we have that
$\widetilde{\mathbf{W}}^{(t+1)}_m$ is also an optimal solution to
the problem (R-LBM).

Evidently, performing the above procedure for at most $r$ times, we
will obtain a rank-1 solution $\mathbf{W}^*_m$ that solves the
problem (LBM). Now the question is that under what condition can we
find $\mathbf{D}^{(t)}$ that satisfies \eqref{eqDH}--\eqref{eqDI} in
each iteration $t$. Note that $\mathbf{D}^{(t)}$ is a $r^{(t)}\times
r^{(t)}$ Hermitian matrix, hence finding $\mathbf{D}^{(t)}$ that
satisfies \eqref{eqDH}--\eqref{eqDI} is equivalent to solving a
system of three linear equations with $(R^{(t)})^2$ unknowns
\footnote{The number of unknowns for the real part of
$\mathbf{D}^{(t)}$ is $\frac{(R^{(t)}+1)R^{(t)}}{2}$, and the number
of unknowns for the imaginary part of $\mathbf{D}^{(t)}$ is
$\frac{(R^{(t)}-1)R^{(t)}}{2}$.}. As long as $(R^{(t)})^2>3$, the
linear system is underdetermined and such $\mathbf{D}^{(t)}$ can be
found. Consequently, the RRP procedure, when terminated, gives us a
$\mathbf{W}^*_m$ with $\rank^2(\mathbf{W}^*_m)\le 3$. As the rank of
a matrix is an integer, we must have $\rank(\mathbf{W}^*_m)=1$. It
is important to note, however, that the ability of the RRP procedure
to recover a rank-1 solution for problem (R-LBM) lies in the fact
that {\it we only have three linear terms of $\mathbf{W}_m$ in both
the objectives and the constraints}. This results in solving a
linear system with {\it three} equations in each iteration of the
RRP procedure. If we have an additional linear constraint of the
form $\trace(\mathbf{B}\mathbf{W}_m)\le c$ for some constant $c$,
the RRP procedure may produce a solution $\mathbf{W}^*_m$ with
$\rank^2(\mathbf{W}^*_m)\le 4$, which does not guarantee
$\rank(\mathbf{W}^*_m)=1$.

We have used the RRP procedure to identify the structure of problem
(R-LBM) that allows for the existence of a rank-1 solution. However
in practice this procedure is not that useful as it requires solving
(R-LBM) to begin with. Therefore we will use our own algorithm
listed in Table \ref{tableUtilityMaximization} to directly get a
rank-1 solution of (R-LBM). Summarizing the above discussion, we
propose the following algorithm, named Successive and Sequential
Convex Approximation Beam Forming (SSCA-BF):

1) {\bf Initialization}: Let $t=0$, randomly choose a set of
feasible covariances $\mathbf{W}_m^{0},~\forall~m\in\mathcal{M}$.

2) {\bf Information Exchange}: Choose $m=M\oslash t$, let each BS
$q\ne m$ compute and transfer $T_q(\mathbf{W}^t)$ to BS $m$.

3) {\bf Maximization}: BS $m$ use the procedure in Table
\ref{tableUtilityMaximization} to obtain a solution
${\mathbf{W}}_m^{t+1}$ of problem (LBM) with the objective function
$U_m(\mathbf{W}_m, {\mathbf{W}}^t_{-m})$. Let
$\mathbf{W}^{t+1}=[\mathbf{W}_m^{t+1}, \mathbf{W}^t_{-m}]$.

%5) {\bf Update}: If $U_{m}({\mathbf{W}}^*_{m},{\mathbf{W}}^t_{-m}
%)\ge U_{m}({\mathbf{W}}^t)$ Set
%$\mathbf{W}^{t+1}=[{\mathbf{W}}_m^{*},\mathbf{W}^{t}_{-m}]$;
%otherwise Set $\mathbf{W}^{t+1}={\mathbf{W}}^{t}$.

4) {\bf Continue}: If
$|R(\mathbf{W}^{t+1})-R(\mathbf{W}^{t+1-M})|<\epsilon$, stop.
Otherwise, set $t=t+1$, go to Step 2).

In Step 4), $\epsilon>0$ is the stopping criteria. The above
algorithm is distributed in the sense that as long as the BS $m$
have the information specified in Step 2) and the channels
$\{\mathbf{H}_{m,q}\}_{q\ne m}$, it can carry out the computation by
itself. %This set of information is exchanged through the backhaul
%network, which in many cases is viewed as a bottleneck for intercell
%cooperation. In section \ref{secSimulation}, we will compare the
%backhaul information exchange needed for different algorithms.
 %the necessary information that is
%needed from all the other BSs $q\ne m$ is: a) the total interference
%at time $t$, $I_q({\mathbf{W}^t_{-q}})$; b) receiving power of the
%useful signal
%$\mathbf{h}^H_{q,q}{\mathbf{W}}^t_{q}\mathbf{h}_{q,q}$.
%Combining this result with the fact that the system sum rate is
%upper bounded, we have the following simple convergence result of
%the SSCA-BF algorithm.
\newtheorem{T1}{Theorem}
\begin{T1}\label{theromConvergenceSingleUser}
{\it The sequence {\small $\{R(\mathbf{W}^t)\}$} produced by the
SSCA-BF algorithm is non-decreasing and converges. Moreover every
limit point of the sequence {\small $\{\mathbf{W}^t\}$} is a
stationary solution to the problem (SRM).}
\end{T1}
\begin{proof}
Fix a iteration $t$ and let $m=M\oslash t$. Due to the fact that we
are able to solve the problem (LBM) exactly, we have
$U_m(\mathbf{W}^{t+1}_{m},{\mathbf{W}}^t_{-m})\ge
U_m({\mathbf{W}}^t_m)$. Using \eqref{eqRateIncreasePerUser} and the
fact that $U_m({\mathbf{W}}^t_m)=R({\mathbf{W}}^t_m)$, we have
{\small
\begin{align}
R({\mathbf{W}}^{t+1})=R({\mathbf{W}}^{t+1}_{m},{\mathbf{W}}^t_{-m})\ge
U_{m}({\mathbf{W}}^{t+1}_{m},{\mathbf{W}}^t_{-m} )\ge
U_{m}({\mathbf{W}}^t_{m},{\mathbf{W}}^t_{-m} )=
R({\mathbf{W}}^t).\label{eqRateIncrease}
\end{align}}
Clearly the system sum rate is upper bounded, then the sequence
$\{R(\mathbf{W}^t)\}_{t=1}^{\infty}$ is nondecreasing and converges.
Take any converging subsequence of $\{\bfW^{t}\}_{t=1}^{\infty}$,
and denote it as $\{\bfW^{l}\}_{l=1}^{\infty}$. Define
$\bfW^*=\lim_{l\to\infty}\bfW^l$. For all BS $m\in\mathcal{M}$, we
must have $U_m(\bfW_m^*,\bfW^*_{-m})\ge
U_m(\bfW_m,\bfW^*_{-m}),~\forall~\bfW_m\in\mathcal{F}_m$, i.e.,
\begin{align}
\bfW^*_m\in\arg\max_{\mathbf{W}_m\in\mathcal{F}_m}
U_m(\bfW_m,\bfW^*_{-m}), \ \forall \ m\in\mathcal{M}.
\end{align}
Checking the KKT conditions of the above $M$ optimization problems,
it is straightforward to see that they are equivalent to the KKT
condition of the original problem (SRM). It follows that $\bfW^*$ is
a KKT point of the problem (SRM). In summary, any limit point of the
sequence $\{\bfW^{t}\}_{t=1}^{\infty}$ is a KKT point of the problem
(SRM).
\end{proof}

%\subsection{Single Cell Network with multiple Users}
%In this section, we consider another simple network configuration in
%which there is a single cell with multiple users, and propose an
%algorithm similar to that in the previous subsection to approach the
%solution of the problem (SRM). In this case, we use $\mathbf{W}_i$
%to denote the covariance intended for user $i$. The user $i$'s
%utility can be simplified as
%\begin{align}
%U_{i}(\mathbf{W}_{i},\widehat{\mathbf{W}}_{-i})&=R(\mathbf{W}_{i},\widehat{\mathbf{W}}_{-i})
%+R_{-i}\left(\widehat{\mathbf{W}}\right)- \sum_{j\ne
%i}Tr\left[t^{i}_{j}\left(\widehat{\mathbf{W}}\right)\mathbf{H}_{j}(\mathbf{W}_{i}-
%\widehat{\mathbf{W}}_{i})\right]\label{eqUtilitySingleBS}
%\end{align}
%where $\mathbf{H}_{i}$ is the channel between the BS and user $i$;
%$t^i_j(.)$ is defined similarly as in \eqref{eqTax} and
%\eqref{eqTaxSingleUser}.

\vspace{-0.1cm}
\section{Multi-cell Network with Multiple Users In Each
Cell}\label{secMultipleUser}

In this section, we consider the network with multiple users per
cell. In this scenario, we can no longer perform the SSCA-BF
algorithm  {\it cyclicly among all the users} to maximize the system
sum rate. The reason is that different users in the same BS share a
{\it coupled constraint} {\small
$\trace(\sum_{i\in\mathcal{N}_m}\mathbf{W}_{m,i})\le\bar{p}_m$}. For
example, consider a network with a single BS $m$ and multiple users.
Suppose at time $0$,
$\mathbf{W}^0_{m,i}=\mathbf{0},~\forall~i\in\mathcal{N}_m$. Suppose
BS $m$ optimizes user $(m,1)$ first (solving problem (LBM) for user
$(m,1)$ with constraints {\small
$\trace(\mathbf{W}_{m,1})+\trace(\sum_{j\ne 1,
j\in\mathcal{N}_m}\mathbf{W}^0_{m,j})\le\bar{p}_m$} and
$\mathbf{W}_{m,i}\succeq 0$). The covariance so obtained has the
form
$\mathbf{W}^*_{m,1}=\bar{p}_m\frac{\mathbf{h}_{m,m_1}\mathbf{h}^H_{m,m_1}}{||\mathbf{h}_{m,m_1}||}$,
and must have the property $\trace(\mathbf{W}^*_{m,1})=\bar{p}_m$.
Then all the subsequent computations ($t=1,\cdots$) within BS $m$
yields $\mathbf{W}^*_{m,i}=\mathbf{0}$, $\forall~i\ne1$, because
each of the problem has to satisfy the joint power constraint.

In order to avoid the above problem, we propose to compute the
covariance matrices {\it BS by BS}, instead of user by user, i.e, to
update the set
$\mathbf{W}_{m}=\{\mathbf{W}_{m,i}\}_{i\in\mathcal{N}_m}$ at the
same time, and cycle through the BSs. To this end, we first identify
a set of {\it per-BS} lower bounds that will be useful in the
subsequent development.
\newtheorem{P3}{Proposition}
\begin{P1}\label{propLowerBoundBS}
{\it For all feasible ${\mathbf{W}}_m$ and a fixed
$\widehat{\mathbf{W}}$ we have the following inequality {\small
\begin{align}
R_{m}(\mathbf{W}_{m},\widehat{\mathbf{W}}_{-m})+R_{-m}\left(\widehat{\mathbf{W}}\right)-
\sum_{i\in\mathcal{N}_m}\sum_{q\ne
m}\sum_{j\in\mathcal{N}_q}\trace\left[T_{q,j}\left(\widehat{\mathbf{W}}\right)\mathbf{H}_{m,q_j}(\mathbf{W}_{m,i}-
\widehat{\mathbf{W}}_{m,i})\right]\le
R(\mathbf{W}_{m},\widehat{\mathbf{W}}_{-m})\label{eqLowerBoundBS}
\end{align}}
where the equality is achieved when
$\mathbf{W}_{m}=\widehat{\mathbf{W}}_{m}$. Define the left hand side
of \eqref{eqLowerBoundBS} as
$\bar{U}_{m}(\mathbf{W}_{m},\widehat{\mathbf{W}}_{-m})$, which is
the lower bound associated with BS $m$.}\end{P1}
\begin{proof}
We can verify, similarly as in Proposition \ref{propConvex}, that
$R_{-m}\left(\mathbf{W}_{m},{\mathbf{W}}_{-m}\right)$ is {\it
jointly convex} with the set of matrices
$\{\mathbf{W}_{m,i}\}_{i\in\mathcal{N}_m}$. Then the lower bound in
\eqref{eqLowerBoundBS} can be obtained by Taylor expansion. Due to
space limit, we do not reiterate the proof here.
\end{proof}
% Similarly as in Section
%\ref{secSingleUser}, the per-BS lower bound $\bar{U}_{m}(.)$ can be
%used by BS $m$ to update its covariance matrices. If BS $m$ can find
%some $\mathbf{W}^*_{m}$ such that {\small
%$\bar{U}_{m}(\mathbf{W}^*_{m},\widehat{\mathbf{W}}_{-m})>
%\bar{U}_{m}(\widehat{\mathbf{W}}_{m},\widehat{\mathbf{W}}_{-m})$},
%which is equivalent to {\small
%\begin{align}
%R_{m}(\mathbf{W}^*_{m},\widehat{\mathbf{W}}_{-m})-
%\sum_{i\in\mathcal{N}_m}\sum_{q\ne
%m}\sum_{j\in\mathcal{N}_q}Tr\left[T_{q,j}\left(\widehat{\mathbf{W}}\right)\mathbf{H}_{m,j}(\mathbf{W}^*_{m,i}-
%\widehat{\mathbf{W}}_{m,i})\right]>
%R_{m}(\widehat{\mathbf{W}}_{m},\widehat{\mathbf{W}}_{-m})\label{eqPropertyLowerBoundBS}
%\end{align}}
%then we must have $R(\mathbf{W}^*_{m},\widehat{\mathbf{W}}_{-m})>
%R(\widehat{\mathbf{W}})$.

Unfortunately, unlike the lower bound $U_{m}(.)$ obtained for the
single user per BS case, $\bar{U}_{m}(.)$ is {\it not} concave in
$\mathbf{W}_{m}$, due to the non-concavity of
{\small$R_m(\mathbf{W}_{m},\mathbf{W}_{-m})$} w.r.t. $\mathbf{W}_m$.
In the following, we propose a heuristic algorithms to optimize the
per-BS lower bound.

We first express the lower bound
$\bar{U}_{m}(\mathbf{W}_{m},\widehat{\mathbf{W}}_{-m})$ in an
equivalent form (where
$\mathbf{w}_m\triangleq\{\mathbf{w}_{m,i}\}_{i\in\mathcal{N}_m}$){\small
\begin{align}
\bar{U}_{m}(\mathbf{w}_{m},\widehat{\mathbf{w}}_{-m})&\triangleq
R_{m}(\mathbf{w}_{m},\widehat{\mathbf{w}}_{-m})+R_{-m}\left(\widehat{\mathbf{w}}\right)-
\sum_{i\in\mathcal{N}_m}\sum_{q\ne
m}\sum_{j\in\mathcal{N}_q}T_{q,j}\left(\widehat{\mathbf{w}}\right)
\left(\mathbf{w}^H_{m,i}\mathbf{H}_{m,q_j}\mathbf{w}_{m,i}-
\widehat{\mathbf{w}}^H_{m,i}\mathbf{H}_{m,q_j}\widehat{\mathbf{w}}_{m,i}\right)\nonumber.
\end{align}}
Then individual BSs' lower bound optimization problem is
\begin{align}
\max_{\mathbf{w}_m}&\quad \bar{U}_{m}(\mathbf{w}_{m},\widehat{\mathbf{w}}_{-m})\label{LBM-BS}\\
\textrm{s.t.}&\quad
\sum_{i\in\mathcal{N}_m}\mathbf{w}^H_{m,i}\mathbf{w}_{m,i}\le\bar{p}_m\nonumber
\end{align}

Take the derivative of the Lagrangian of the problem \eqref{LBM-BS}
w.r.t. $\mathbf{w}_{m,i}$ to be zero, we obtain
%\begin{align}
%\frac{2/\ln(2)\times\mathbf{H}_{m,i}\mathbf{w}_{m,i}}{\sum_{q\ne
%m}\sum_{j\in\mathcal{N}_q}\widehat{\mathbf{w}}^H_{q,j}\mathbf{H}_{q,i}\widehat{\mathbf{w}}_{q,j}+\sum_{l\in\mathcal{N}_m
%}{\mathbf{w}}^H_{m,l}\mathbf{H}_{m,i}{\mathbf{w}}_{m,l}}-2\sum_{(q,j)\ne(m,i)}T_{q,j}(\widehat{\mathbf{w}})
%\mathbf{H}_{m,j}\mathbf{w}_{m,i}-2\lambda\mathbf{w}_{m,i}
%\end{align}
{\small\begin{align} &\ln(2)\left(\sum_{q\ne
m}\sum_{j\in\mathcal{N}_q}T_{q,j}(\widehat{\mathbf{w}}_q,
\widehat{\mathbf{w}}_{-q}) \mathbf{H}_{m,q_j}+\sum_{l\ne
i,l\in\mathcal{N}_m}T_{m,l}(\mathbf{w}_m,\widehat{\mathbf{w}}_{-m})
\mathbf{H}_{m,m_l}+\mu_m\mathbf{I}_{p}\right)\mathbf{w}_{m,i}\nonumber\\
&= \frac{\mathbf{H}_{m,m_i}\mathbf{w}_{m,i}}{\sum_{q\ne
m}\sum_{j\in\mathcal{N}_q}\widehat{\mathbf{w}}^H_{q,j}\mathbf{H}_{q,m_i}\widehat{\mathbf{w}}_{q,j}+
\sum_{l\in\mathcal{N}_m
}{\mathbf{w}}^H_{m,l}\mathbf{H}_{m,m_i}{\mathbf{w}}_{m,l}},~\forall~i\in\mathcal{N}_m\label{eqLagrangianZero}
\end{align}}
where $\mu_m\ge 0$ is the dual variable associated with the power
constraint, and
$T_{m,l}\left(\mathbf{w}_m,\widehat{\mathbf{w}}_{-m}\right)$ is
defined {\small
\begin{align}
T_{m,l}\left(\mathbf{w}_m,\widehat{\mathbf{w}}_{-m}\right)&=
\frac{1/\ln(2)}{\sum_{q\ne m,
j\in\mathcal{N}_q}\widehat{\mathbf{w}}^H_{q,j}\mathbf{H}_{q,m_l}\widehat{\mathbf{w}}_{q,j}+
\sum_{i\in\mathcal{N}_m
}{\mathbf{w}}^H_{m,i}\mathbf{H}_{m,m_l}{\mathbf{w}}_{m,i}}\times\nonumber\\
&\frac{\mathbf{w}^H_{m,m_l}\mathbf{H}_{m,l}{\mathbf{w}}_{m,l}}
{\sum_{q\ne m,
j\in\mathcal{N}_q}\widehat{\mathbf{w}}^H_{q,j}\mathbf{H}_{q,m_l}\widehat{\mathbf{w}}_{q,j}+\sum_{i\ne
l, i\in\mathcal{N}_m
}{\mathbf{w}}^H_{m,i}\mathbf{H}_{m,m_l}{\mathbf{w}}_{m,i}}.
\end{align}}
A tuple $(\mu_m, \mathbf{w}_i)$ that satisfies the $N$ equations in
\eqref{eqLagrangianZero} as well as the complementarity and
feasibility conditions $\mu_m\ge 0,
\mu_m(\bar{p}_m-\sum_{i\in\mathcal{N}_m}\mathbf{w}^H_{m,i}\mathbf{w}_{m,i})=
0$ and
$\bar{p}_m-\sum_{i\in\mathcal{N}_m}\mathbf{w}^H_{m,i}\mathbf{w}_{m,i}\ge
0$ is a stationary solution to the problem \eqref{LBM-BS}. Let us
define{\small
\begin{align}
\mathbf{M}_{m,i}(\mu_m, \widehat{\mathbf{w}})\triangleq
\ln(2)\left(\sum_{(q,j)\ne (m,i)}T_{q,j}(\widehat{\mathbf{w}}_m,
\widehat{\mathbf{w}}_{-m})
\mathbf{H}_{m,q_j}+\mu_m\mathbf{I}_{p}\right).
\end{align}}
 %{\small
%\begin{align}
%\mathbf{M}_{m,i}(\lambda, \widehat{\mathbf{w}})&\triangleq
%\ln(2)\left(\sum_{(q,j)\ne (m,i)}T_{q,j}(\widehat{\mathbf{w}})
%\mathbf{H}_{m,j}+\lambda\mathbf{I}_{p}\right)\nonumber\\
%\mathbf{L}_{m,i}(\widehat{\mathbf{w}})&\triangleq
%\frac{\mathbf{H}_{m,i}}{\sum_{q\ne
%m}\sum_{j\in\mathcal{N}_q}\widehat{\mathbf{w}}^H_{q,j}\mathbf{H}_{q,i}\widehat{\mathbf{w}}_{q,j}+\sum_{l\in\mathcal{N}_m
%}{\widehat{\mathbf{w}}}^H_{m,l}\mathbf{H}_{m,i}{\widehat{\mathbf{w}}}_{m,l}},~\forall~i\in\mathcal{N}_m.\label{eqWUpdate}
%\end{align}}
It is shown in \cite[Proposition 1]{venturino10} that the optimal
beam vector $\mathbf{w}_{m,i}$ that satisfy \eqref{eqLagrangianZero}
must satisfy the following identity
\begin{align}
\mathbf{w}_{m,i}=\beta_{m,i}(\mu_m)\mathbf{M}^\dag_{m,i}(\mu_m,
\widehat{\mathbf{w}})\mathbf{h}_{m,m_i}\label{eqWUpdate}
\end{align}
for some constant $\beta_{m,i}(\mu_m)$ that can be computed
as{\small
\begin{align}
\beta_{m,i}(\mu_m)=\sqrt{
\frac{\left[\mathbf{h}^H_{m,m_i}\mathbf{M}^{\dag}_{m,i}(\mu_m,
\widehat{\mathbf{w}})\mathbf{h}_{m,m_i}-I_{m,i}(\widehat{\mathbf{w}}_{-(m,i)})\right]^+}
{(\mathbf{h}^H_{m,m_i}\mathbf{M}^{\dag}_{m,i}(\mu_m,
\widehat{\mathbf{w}})\mathbf{h}_{m,m_i})^2}}\label{eqBeta}.
\end{align}}
As a result, we can compute
$\{\mathbf{w}_{m,i}\}_{i\in\mathcal{N}_m}$ by first computing
$\beta_{m,i}(\mu_m)$ according to \eqref{eqBeta},
%carrying out the following computation
%\begin{align}
%\mathbf{w}_{m,i}=\mathbf{M}^\dag_{m,i}(\lambda,
%\widehat{\mathbf{w}})\mathbf{L}_{m,i}(\widehat{\mathbf{w}})\widehat{\mathbf{w}}_{m,i},~
%\forall~i\in\mathcal{N}_m\label{eqWUpdate}
%\end{align}
and then use bisection (similarly as in the classic water filling
algorithm) to find an appropriate $\mu_m\ge 0$ such that the power
constraint for BS $m$ is satisfied. To this end, we propose a
Sequential Beamforming (S-BF) algorithm:

1) {\bf Initialization}: Let $t=0$, randomly choose a set of
feasible transmission beams
$\mathbf{w}_m^{0},~\forall~m\in\mathcal{M}$.

2) {\bf Information Exchange}: Choose $m=\{(t+1) \textrm{mode}(M)
\}+1$, let each BS $q\ne m$ compute and transfer
$\{T_{q,j}(\mathbf{w}^t)\}_{j\in\mathcal{N}_q}$ to BS $m$ through
the backhaul network.

3) {\bf Computation}: BS $m$ updates its beam vectors according to
\eqref{eqWUpdate} and \eqref{eqBeta}, with
$\widehat{\mathbf{w}}={\mathbf{w}}^t$. Use bisection to find $\mu_m$
that ensures the power constraint. Obtain the solution
${\mathbf{w}}_{m}^{*}$.

4) {\bf Update}: If
$\bar{U}_{m}({\mathbf{w}}^*_{m},{\mathbf{w}}^t_{-m} )\ge
\bar{U}_{m}({\mathbf{w}}^t)$ Set
$\mathbf{w}^{t+1}=[{\mathbf{w}}_m^{*},\mathbf{w}^{t}_{-m}]$;
otherwise Set $\mathbf{w}^{t+1}={\mathbf{w}}^{t}$.

5) {\bf Continue}: If $|R(\mathbf{w
}^{t+1})-R(\mathbf{w}^{t+1-M})|<\epsilon$, stop. Otherwise, set
$t=t+1$, go to Step 2).

Note that in Step 4) we check if the lower bound is increased. If
this is indeed the case, we accept the new set of beams
$\mathbf{w}^*_m$. This procedure ensures $R(\mathbf{w}^{t+1})\ge
R(\mathbf{w}^t)$.

The S-BF algorithm is a variant/extention of the the ICBF algorithm
proposed in \cite{venturino10}: Step 2) and Step 3) of S-BF is
 a sequential version of the ICBF algorithm.
However, the S-BF algorithm does have several advantages/differences
to the ICBF algorithm: \emph{i)} The ICBF
 tries to solve the KKT system of the problem (SRM), while
S-BF tries to optimize the per-BS lower bound {\it for each BS};
\emph{ii)} In S-BF algorithm the BSs update sequentially while in
the ICBF algorithm the BSs update at the same time. One important
consequence of such difference in updating schedule is the amount of
information exchange needed in each iteration: in our algorithm, all
BSs only need to send a single copy of their local information to a
{\it single} BS, while in ICBF algorithm, they need to send to {\it
all other} BSs. As will be shown in Section \ref{secSimulation}, the
total information exchange needed for both S-BF and SSCA-BF
algorithm is significantly less than the ICBF algorithm; \emph{iii)}
Due to the utilization of the per-BS lower bound in Step 4), the
system sum rate of the proposed S-BF algorithm monotonically
increases and converges, while the ICBF algorithm does not possess
such convergence guarantee; \emph{iv)} In S-BF algorithm, there is
no ``inner iteration", in which all the BSs update their beam
vectors at the same time to reach some {\it intermediate
convergence} (note that in ICBF algorithm, the convergence of the
inner iteration is {\it not} guaranteed). Such ``inner iteration" is
undesirable, because \emph{a)} it is hard to decide on, in a
distributed fashion, whether convergence has been reached and \emph
b) in each of such inner iterations, extra feedback information
needs to be exchanged between the BSs and their users.

\vspace{-0.3cm}
\section{Numerical Results}\label{secSimulation}
\begin{figure*}[htb] \vspace*{-.5cm}
    \begin{minipage}[t]{0.49\linewidth}
    \centering
    {\includegraphics[width=
1\linewidth]{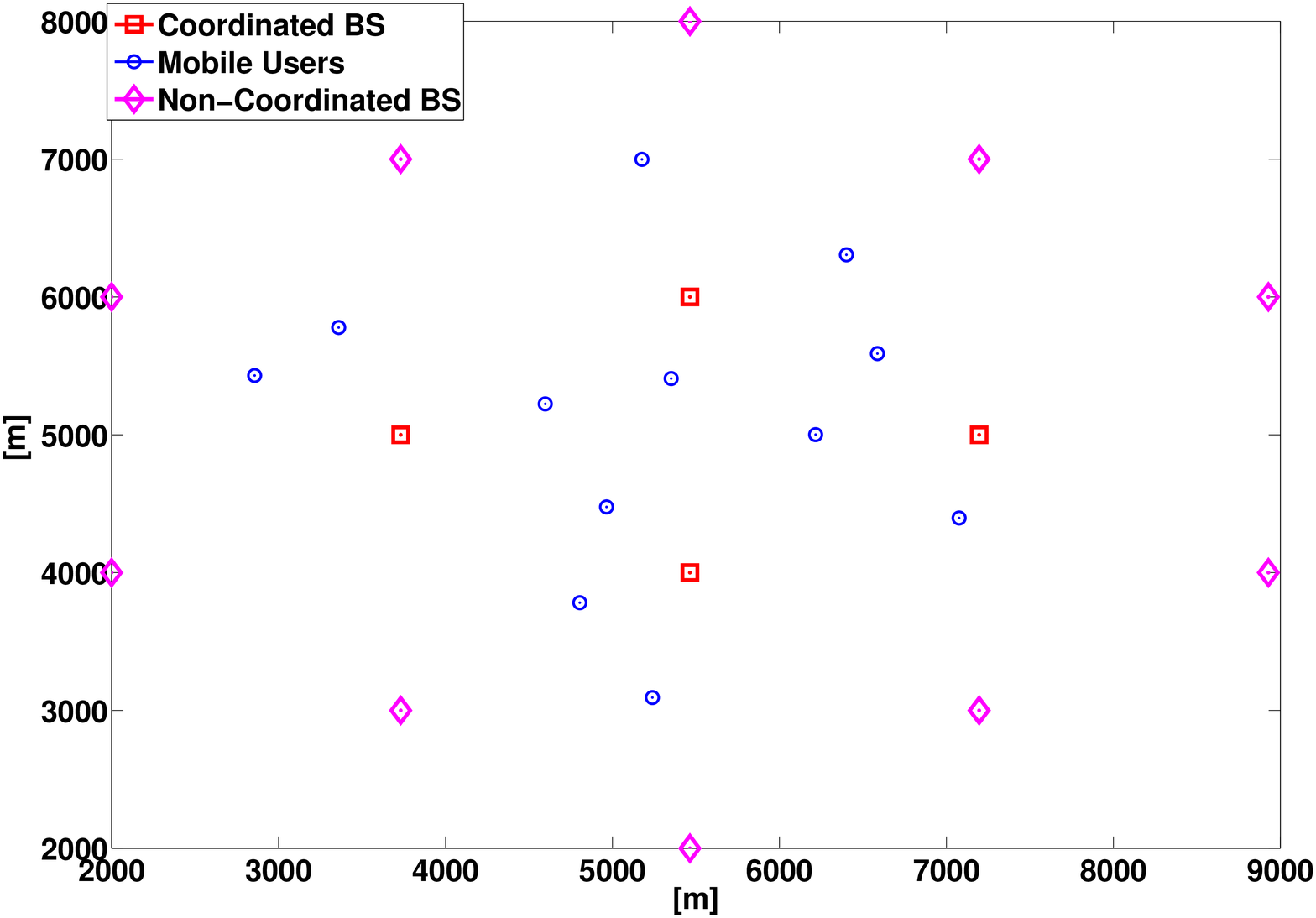} \vspace*{-0.8cm}\caption{Topology of
simulated network.}\label{figTopology} \vspace*{-0.1cm}}
\end{minipage}\hfill
    \begin{minipage}[t]{0.49\linewidth}
    \centering
    {\includegraphics[width=
1\linewidth]{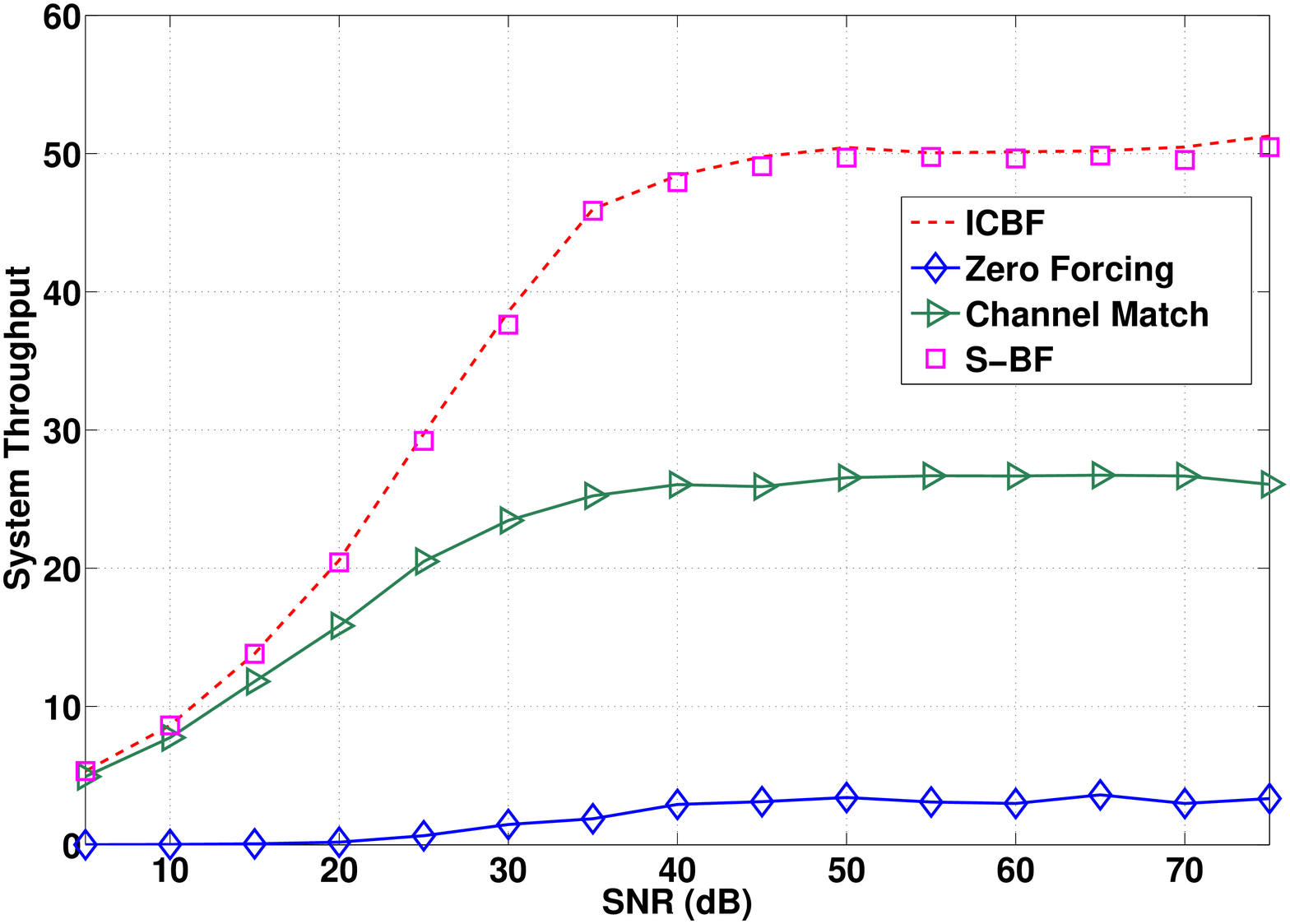}
\vspace*{-0.8cm}\caption{Comparison of system throughput of
Different Algorithms. $K=5$, $N=5$, $M=4$. Users $i\in\mathcal{N}_m$
uniformly placed within $d_{m,m_i}\in[200,~1000]$ meters within each
BS.}\label{figSumRate1} \vspace*{-0.1cm}}
\end{minipage}
\vspace*{-0.4cm}
    \end{figure*}

In this section, we give numerical results demonstrating the
performance of the proposed algorithms. We mainly consider a network
with a set $\mathcal{W}$ of BS, where $|\mathcal{W}|=14$ (see Fig.
\ref{figTopology} for the system topology of the network with
randomly generated user locations). $4$ of the BSs are coordinated
for transmission (in the set $\mathcal{M}$), i.e., $M=4$. All other
BSs' (in the set $\mathcal{W}/\mathcal{M}$) transmission is regarded
as noise. The BS to BS distance is 2 km. Let $d_{q,m_i}$ be the
distance between BS $q$ and $i$th user in $m$th cell.
 The channel coefficients are modeled as zero mean circularly
symmetric complex Gaussian vector with
$\left({200}/{d_{q,m_i}}\right)^{3.5}L_{q,m_i}$ as variance for each
part, where $10\log10(L_{q,m_i})$ is a real Gaussian random variable
modeling the shadowing effect with zero mean and standard deviation
8. The environmental noise power is modeled as the power of thermal
noise plus the power of noises/interferences generated by
non-coordinating BSs: $
c_{m,i}=\sigma^2+\sum_{w\in\mathcal{W}-\mathcal{M}}\left({200}/{d_{w,m_i}}\right)^{3.5}L_{w,m_i}\bar{p}_w$.
We take $\bar{p}_m=1$ for all $m\in\mathcal{W}$, and define the
$SNR$ as $10\log10(\bar{p}_m/\sigma^2)$. The stopping criteria is
set to be $\epsilon=10^2$ for all the algorithms.

\begin{figure*}[htb] \vspace*{-.5cm}
    \begin{minipage}[t]{0.49\linewidth}
    \centering
    {\includegraphics[width=
1\linewidth]{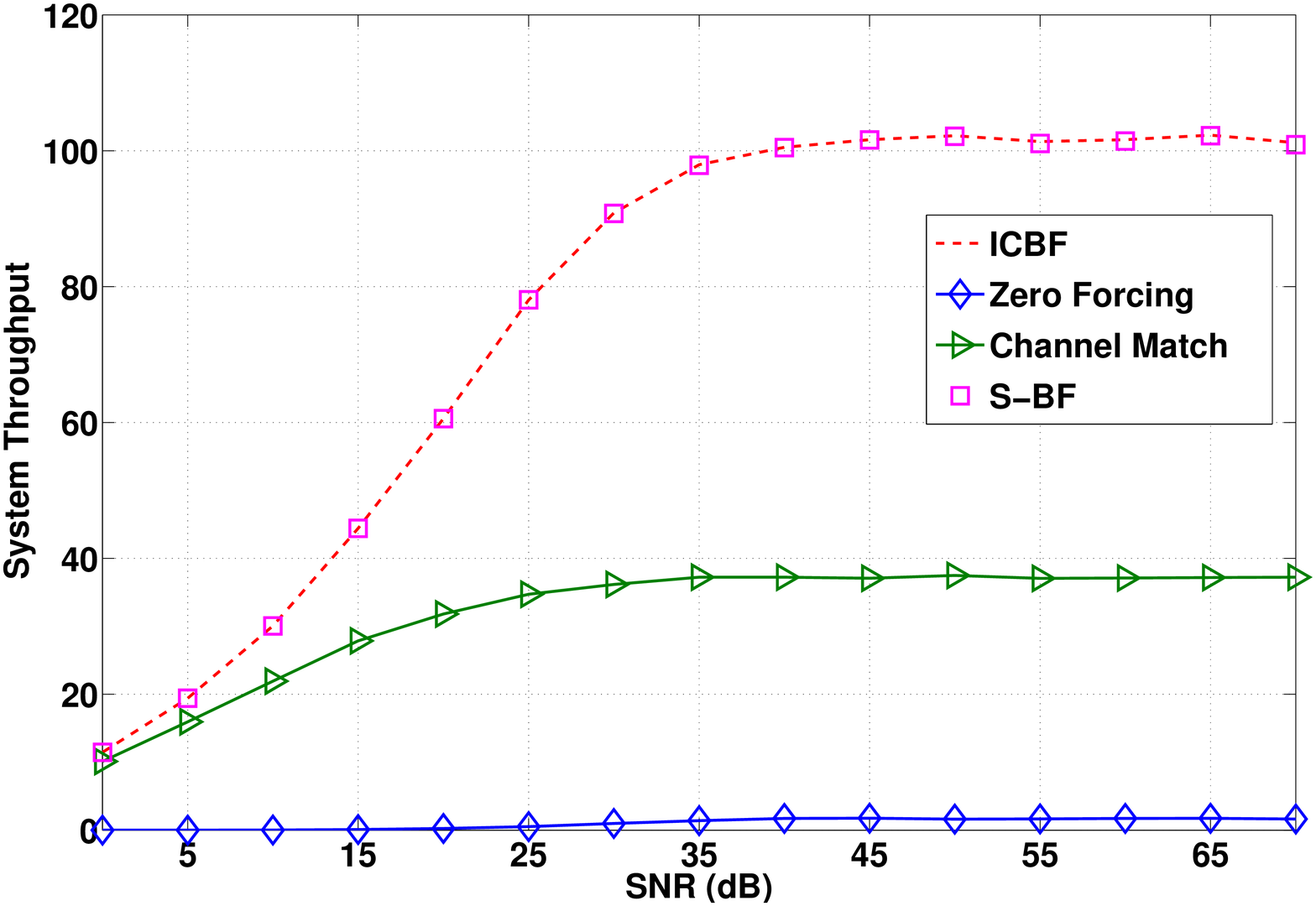}
\vspace*{-0.6cm}\caption{Comparison of system throughput of
Different Algorithms. $K=10$, $N=10$, $M=4$. Users
$i\in\mathcal{N}_m$ uniformly placed within
$d_{m,m_i}\in[200,~1000]$ meters within each BS.}\label{figSumRate2}
\vspace*{-0.1cm}}
\end{minipage}\hfill
    \begin{minipage}[t]{0.49\linewidth}
    \centering
    {\includegraphics[width=
1\linewidth]{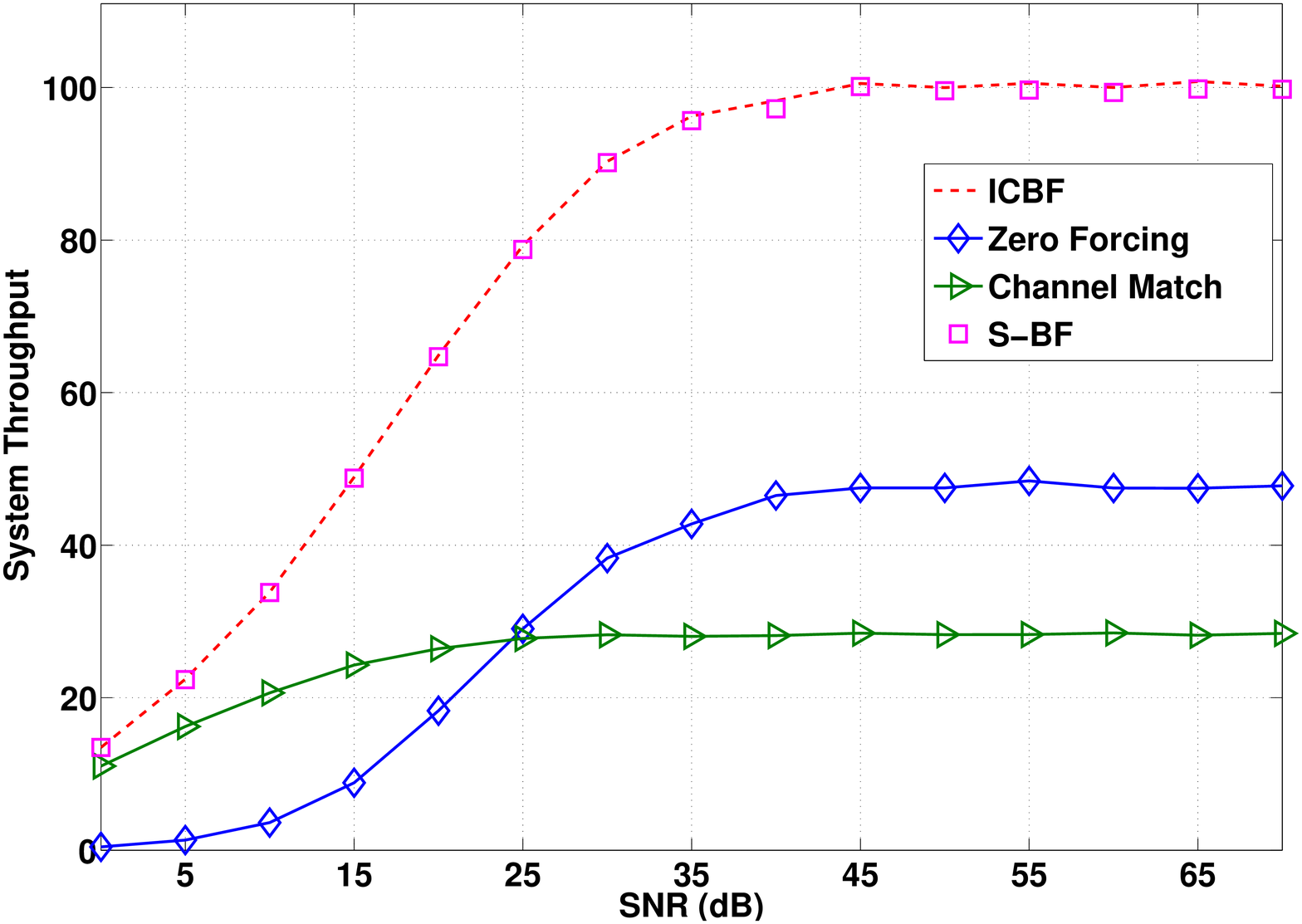}
\vspace*{-0.6cm}\caption{Comparison of system throughput of
different Algorithms. $K=5$, $N=5$, $M=4$. Users $i\in\mathcal{N}_m$
uniformly placed within $d_{m,m_i}\in[200,~300]$ meters within each
BS.}\label{figSumRate3} \vspace*{-0.1cm}}
\end{minipage}
\vspace*{-0.5cm}
    \end{figure*}

In Fig. \ref{figSumRate1} and Fig. \ref{figSumRate2}, we consider
networks with $N=K=5$ and $N=K=10$, where the the users
$i\in\mathcal{N}_m$ that are associated with BS $m$ are uniformly
placed within $d_{m,m_i}\in[200,~1000]$ meters. We show the sum rate
performance of the S-BF algorithm comparing with the ICBF algorithm
in \cite{venturino10} and the non-coordinating schemes where the BSs
individually perform zero forcing beamforming and channel matched
filter beamforming. In Fig. \ref{figSumRate3} we consider network
with $N=K=5$ and $d_{m,m_i}\in[200,~300],~\forall~m, i$. Clearly all
the coordinated schemes achieve similar throughput performance,
which is significantly higher than the non-coordinated
schemes. %W%e also observe that the advantage of the coordinated
%scheme decreases when users are concentrated near the BSs (Fig.
%\ref{figSumRate3}). This result is expected because when the number
%of cell edge users decreases, the intercell interference is also
%reduced.

\begin{figure*}[htb] \vspace*{-.5cm}
    \begin{minipage}[t]{0.49\linewidth}
    \centering
    {\includegraphics[width=
1\linewidth]{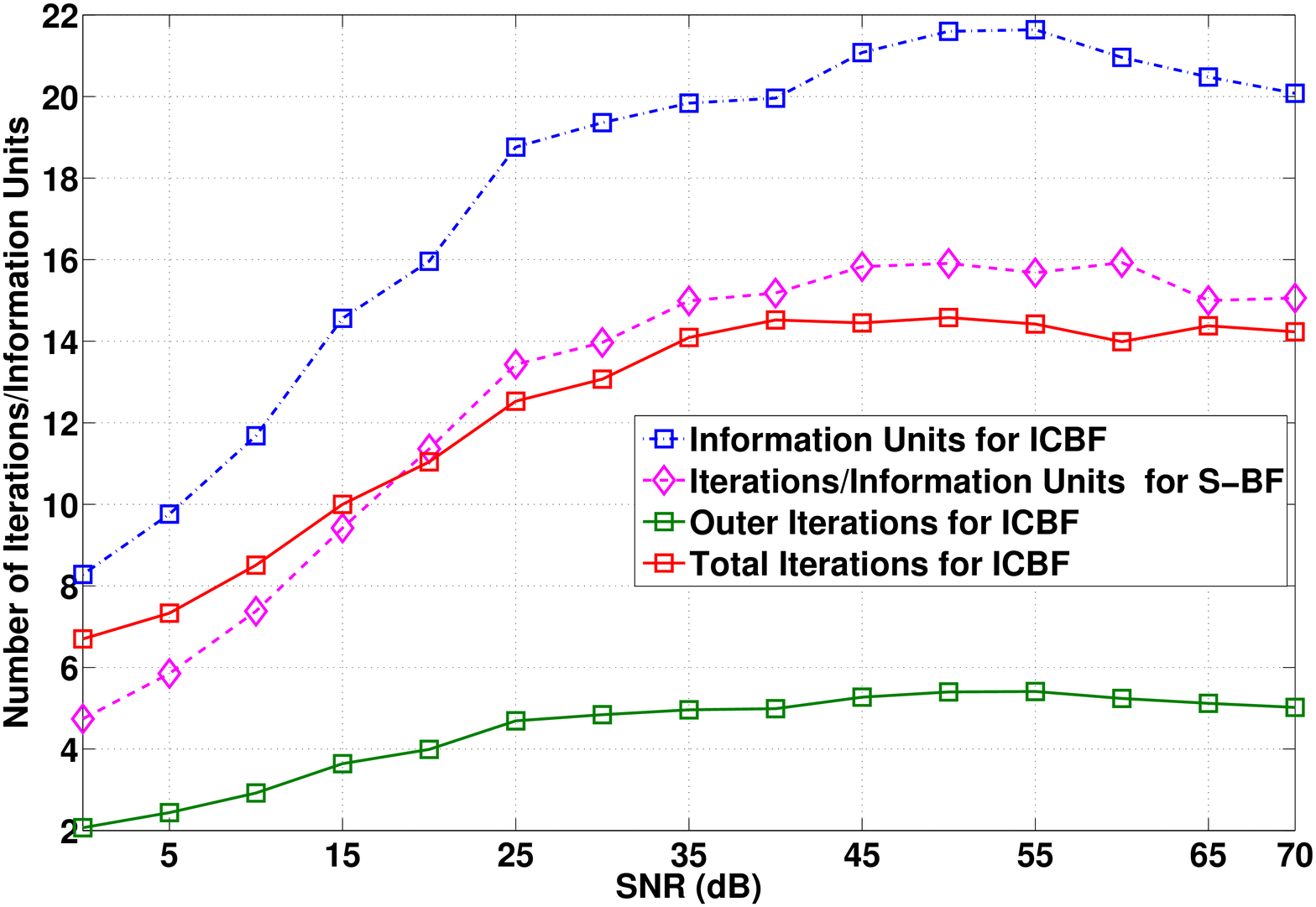}
\vspace*{-0.7cm}\caption{Comparison of the Number of
Iterations/Information Units Needed for Convergence. $K=5$, $N=5$,
$M=4$.}\label{figTimeConvergence1} \vspace*{-0.6cm}}
\end{minipage}\hfill
    \begin{minipage}[t]{0.49\linewidth}
    \centering
    {\includegraphics[width=
1\linewidth]{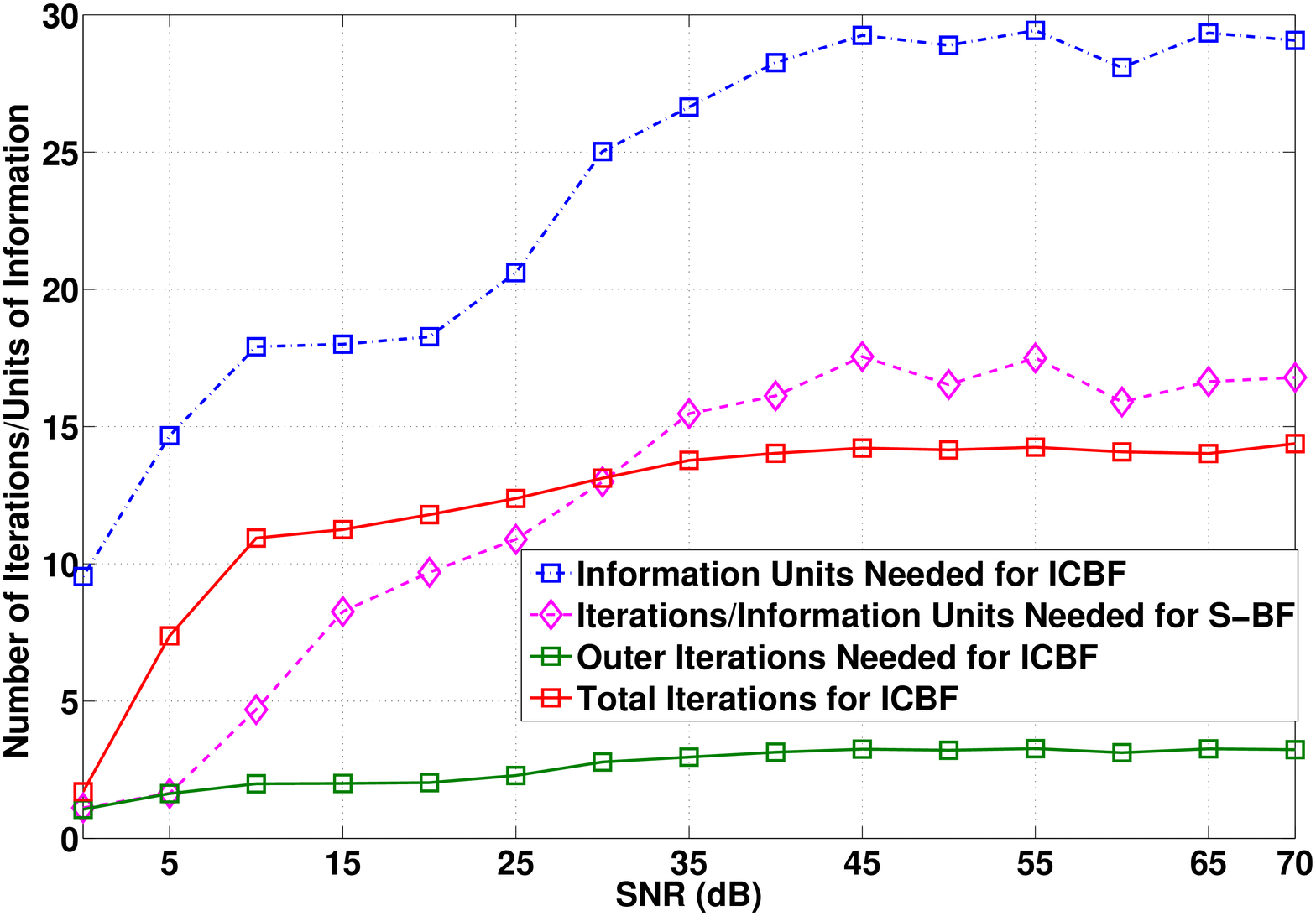}
\vspace*{-0.7cm}\caption{Comparison of the Number of
Iterations/Information Units Needed for Convergence. $K=5$, $N=5$,
$M=9$.}\label{figTimeConvergence2} \vspace*{-0.6cm}}
\end{minipage}
\vspace{-0.6cm}
    \end{figure*}

We then compare the amount of inter-cell information needed for
different coordinated schemes. We define the {\it unit of
information transfer} as the total information needed from the set
of coordinated BS for updating the beam vectors for {\it a single
BS} $m\in\mathcal{M}$. Clearly, in each iteration of the S-BF
algorithm, a single unit of information is needed to go through the
backhaul network, while in ICBF algorithm, $M$ units of information
are needed. In Fig. \ref{figTimeConvergence1} and Fig.
\ref{figTimeConvergence2}, we demonstrate the averaged number of
iterations and the averaged total units of information needed for
different coordinated schemes until convergence. We observe that the
total units of information needed for the proposed SSCA-BF and S-BF
algorithms are around $25\%$ less than the ICBF algorithm when
$M=4$, and around $40\%$ less when $M=9$. \footnote{The network with
$M=9$ is generated similarly as the case of $M=4$, i.e., the center
$9$ BSs are coordinating, while the other BSs around them are
non-coordinating and their transmissions are considered as noises.}
We also emphasize that typically, several {\it inner iterations} are
needed per outer iteration of ICBF, and we have not count the extra
information needed between the BSs and the users in these inner
iterations. As a results, in Fig. \ref{figTimeConvergence1} and Fig.
\ref{figTimeConvergence2} we see that the {\it total iterations}
needed for ICBF algorithm are close to the S-BF algorithm. In all
the simulations presented above, the results are obtained by
averaging over $500$ randomly generated user locations and channel
realizations.

\vspace{-0.2cm}
\section{Conclusion}\label{secConclusion}
In this correspondence, we study the sum rate maximization problem
using beamforming in a multi-cell MISO network. We have explored the
structure of the problem and identified a set of lower bounds for
the system sum rate. For the case of a single user per cell, we
proposed an algorithm that reaches the KKT point of the sum rate
maximization problem. For the case of multiple users per cell, we
propose and algorithm that achieve high system throughput with
reduced backhaul information exchange among the BSs.

\vspace{-0.3cm}
\bibliographystyle{IEEEbib}

\bibliography{ref}

\end{document}